\documentstyle[12pt]{article}
  \advance\oddsidemargin  by -1in
  \advance\evensidemargin by -1in
  \advance\topmargin      by -0.5in
\makeindex

\def\Pom{{\bf I\!P}}
\def\lsim{\mathrel{\rlap{\lower4pt\hbox{\hskip1pt$\sim$}}
    \raise1pt\hbox{$<$}}}         
\def\gsim{\mathrel{\rlap{\lower4pt\hbox{\hskip1pt$\sim$}}
    \raise1pt\hbox{$>$}}}         
\def\beq{\begin{equation}}
\def\eeq{\end{equation}}
\def\bea{\begin{eqnarray}}
\def\eea{\end{eqnarray}}

\begin{document}

\begin{flushright}
{\em ITEP-PH-6/98\\
FZ-IKP(TH)-1998-32}
\end{flushright}
\vspace{1.0cm}

\begin{center}
{\Large \bf The running  BFKL: resolution of  Caldwell's puzzle\\

\vspace{1.0cm}}
{\large \bf N.N. Nikolaev$^{\alpha}$ and  V. R. Zoller$^{\beta}$}\\
\vspace{0.5cm}
$^{\alpha}${ \em
Institut  f\"ur Kernphysik, Forschungszentrum J\"ulich,\\
D-52425 J\"ulich, Germany\\
E-mail: kph154@ikp301.ikp.kfa-juelich.de}\\

$^{\beta}${\em Institute for  Theoretical and Experimental Physics,\\
Moscow 117218, Russia\\
E-mail: zoller@heron.itep.ru}\\
\vspace{0.5cm}
{\bf Abstract}
\end{center}
The HERA data on the proton structure function,  $F_2(x,Q^2)$,
  at very small $x$ and $Q^2$ show  the dramatic departure
of the  logarithmic slope,
  $\partial F_2/\partial\log Q^2$,
 from theoretical predictions based on
 the DGLAP  evolution. We show that the  running BFKL approach
provides 
 the quantitative explanation for
the observed $x$ and/or $Q^2$ -dependence
of  $\partial F_2/\partial\log Q^2$.  \\

\vspace{0.75cm}
Caldwell's presentation 
 of the HERA data in terms of the logarithmic
 derivative $\partial F_2/\partial\log Q^2$
for
the  proton structure function (SF) $F_2(x,Q^2)$
 exhibits the  turn-over of the slope towards small $x$ and/or $Q^2$
 up to currently attainable  $x\sim 10^{-6}$ and 
$Q^2\sim 0.1\,{\rm GeV^2}$ \cite{DERLOG,DERLOG1}.
The  DGLAP-evolution \cite{DGLAP} with GRV  input  \cite{GRV}
 predicts a steady increase
of the derivative
\beq
{\partial F_T^{DGLAP}\over \partial\log Q^2}\propto
 \alpha_S(Q^2)G^{DGLAP}(x,Q^2)
\label{DGLAP}
\eeq
with $1/x$, due to the growth
of the gluon structure function  $G^{DGLAP}(x,Q^2)=xg^{DGLAP}(x,Q^2)$.
 A slight systematic discrepancy of the DGLAP analysis with  small-$x$ data on 
 $\partial F_2/\partial\log Q^2$ 
was found also in \cite{DICK}. 

 The turn-over
point located at $x\sim 5\cdot 10^{-4}$ and $Q^2\sim 5\,{\rm GeV^2}$, in
a commonly believed legitimate pQCD domain.
So, the phenomenon occurs on the interface between "soft" and "hard" physics.
Its explanation within the color dipole  approach 
is based on two observations{\footnote
 {The preliminary results have been reported at the DIS'98 Workshop
\cite{DIS98}}}:\\
i)  specific smallness of the $\log Q^2$-derivative of  sub-leading
terms of the BFKL-Regge expansion for  $F_2$
 at the turn-over point, which is due to the nodal structure of the
 running BFKL eigen-SF's;\\
ii) significant contribution to the small-$Q^2$ proton SF 
 coming from   the non-perturbative component  of the BFKL pomeron.\\

 The $s$-channel
approach to the BFKL equation \cite{BFKL} was developed in terms of 
the color dipole cross section $\sigma(x,r)$ \cite{PISMA1,NZZJETP} 
(hereafter  $\vec r$ is the color dipole moment).
The positive feature of the color dipole picture,
 to be referred to as the running BFKL approach, is 
  consistent incorporating the two crucial properties
of QCD: i) asymptotic freedom (AF), {\sl i.e.},  the running
QCD coupling $\alpha_{S}(r)$ and, ii) the finite
propagation radius $R_{c}$ of perturbative gluons.

The BFKL equation for the interaction cross section
$\sigma(x,r)$ of the color dipole $\vec r$ with the target reads
\bea
{\partial \sigma(x,r) \over \partial \log(1/x)} ={\cal K}\otimes
\sigma(x,r)=~~~~~~~~~~~~~~~~~~~~~~~~~~~~~~~~~~~~~~~~
\nonumber\\ {3 \over 8\pi^{3}} \int d^{2}\vec{\rho}_{1}\,\,
|\vec{{\cal E}}(\vec{\rho}_{1}) - \vec{{\cal E}}(\vec{\rho}_{2})|^{2}
[\sigma(x,\rho_{1})+
\sigma(x,\rho_{2})-\sigma(x,r)]   \, \, .
\label{eq:gBFKL}
\eea
Here the kernel ${\cal K}$ is related to the wave function squared
of the color-singlet $q\bar{q}g$ state with the Weizs\"acker-Williams
(WW) soft gluon.
The quantity
\beq
\vec{{\cal E}}(\vec{\rho})= 
-g_{S}(\rho) \vec{\nabla}_{\rho}
K_{0}(\mu_{G}\rho)=
g_{S}(\rho)\mu_{G}K_{1}(\mu_{G}\rho)
{\vec{\rho}/ \rho}\, 
,
\label{eq:ERHO}
\eeq
 where $R_{c}=1/\mu_{G}$ and $K_{\nu}(x)$ is the modified Bessel
function, describes a Yukawa screened transverse chromoelectric field
of the relativistic quark and
$|\vec{{\cal E}}(\vec{\rho}_{1}) - \vec{{\cal E}}(\vec{\rho}_{2})|^{2}$
%
%
describes the flux (the modulus of the Poynting vector) of WW gluons in
the $q\bar{q}g$ state in which $\vec{r}$ is the $\bar{q}$-$q$ separation
and $\vec{\rho}_{1,2}$ are the $q$-$g$ and $\bar{q}$-$g$ separations
in the two-dimensional impact parameter plane. Our numerical results
 are for the Yukawa screening radius  $R_c= 0.27\,{\rm fm}$.
The recent fits  to the lattice QCD data on
 the field strength correlators suggest similar $R_c$ \cite{MEGGI}.  

 The asymptotic freedom of QCD uniquely
prescribes the chromoelectric field be computed with the running QCD charge
$g_{S}(r)=\sqrt{4\pi \alpha_{S}(r)}$ taken at the
shortest relevant distance, $R_{i}={\rm min}\{r,\rho_{i}\}$ in the
$q\bar{q}g$ system. Although, the so  introduced  running coupling does 
not necessarily exhaust all NLO effects, it correctly describes the crucial 
enhancement of long distance,  and suppression of short distance, effects by AF.

The  properties of  the running color dipole BFKL equation
  responsible for the observed $Q^2$ dependence of 
$\partial F_2/\partial\log Q^2$ are as follows \cite{DIS97,JETP}.
The spectrum of the running BFKL equation is  a series of
 moving poles  in the complex $j$-plane with eigen-functions
\beq
\sigma_{n}(x,r)=\sigma_{n}(r)\exp\left[\Delta_{n}\log(1/x)\right]
\label{eq:SIGRED}
\eeq
 being a solution of
\beq
{\cal K}\otimes \sigma_{n}=\Delta_{n}\sigma_{n}(r).
\label{eq:KO}
\eeq
  The leading eigen-function $\sigma_0(r)$
 is node free.  The  sub-leading  $\sigma_n(r)$ has $n$ nodes.
 The intercepts $\Delta_n$  closely, to better than $10\%$,
 follow the law $ \Delta_n= {\Delta_0/ (n+1)}$ suggested earlier by 
Lipatov \cite{LIPAT86}. The intercept of the leading pole trajectory,
with the above specific choice of $R_c$, 
is $\Delta_0\equiv\Delta_{\Pom}=0.4$.
The sub-leading eigen-functions $\sigma_{n}$  \cite{DIS97,JETP}
 are  very close  to Lipatov's quasi-classical
 solutions \cite{LIPAT86} for $n\gg 1$.
For our specific choice of the infrared regulator, $R_c$,
 the  node of $\sigma_{1}(r)$ is located at 
$r=r_1\simeq 0.05-0.06\,{\rm fm}$, for larger $n$ the first node
moves to a somewhat  larger $r\sim 0.1\, {\rm fm}$.

The color dipole factorization \cite{NZ91}
in conjunction with the explicit form
of the $q\bar q$ light-cone wave function, $\Psi_{qq}(z,r)$, 
 relates the dipole  cross sections $\sigma_{n}(r)$ with the eigen-SF, $f_n(Q^2)$,
\beq
f_n(Q^2)=
{Q^2\over {4\pi\alpha_{em}}}\sum_{q=u,d,c,s}\int_0^1 dz\int 
d^2\vec r|\Psi_{q q}(z,r)|^2\sigma_n(r)\,,
\label{eq:F2CC}
\eeq

 The BFKL-Regge  expansion 
\beq
\sigma(x,r)=\sigma_0(r)(x_0/x)^{\Delta_0}+\sigma_1(r)(x_0/x)^{\Delta_1}+
\sigma_2(r)(x_0/x)^{\Delta_2}+...\,.
\label{eq:sigma}
\eeq
gives the BFKL-Regge expansion for the SF 
\beq
F_2(x,Q^2)=\sum_n f_n(Q^2)(x_0/x)^{\Delta_n}\,.
\label{eq:F2BFKL}
\eeq

The remarkable finding of \cite{PLNZ2,DIS97,JETP} is a good description
 of the HERA data
on the proton SF starting with the Born two-gluon cross section
  $\sigma_B(r)$ as a boundary condition for the running BFKL equation 
(\ref{eq:gBFKL}) at $x_0=0.03$. With such a boundary condition, 
which could well be excessively restrictive, the expansion (\ref{eq:sigma})
fixes uniquely the normalization of the eigen-FS's.
 
The Bjorken variable $x=Q^2/2m_p\nu$ is commonly 
being used for the presentation 
of the experimental 
data even at $Q^2\lsim m^2_{\rho}$, way beyond the kinematical region
 $Q^2\gg m^2_{\rho}$ it has originally been devised for. At small
$Q^2$, the relevant Regge parameter is $2m_p\nu/(Q^2+m^2_{\rho})$ rather than 
the $1/x$. Consequently, in the small-$Q^2$ region the Regge parameter
$x_0/x$ in eqs.(\ref{eq:sigma}) and (\ref{eq:F2BFKL}) must be substituted
by $(x_0/x)(1+m^2_{\rho}/Q^2)\,.$

One more remark on kinematics is in order.
The BFKL-Regge expansion (\ref{eq:F2BFKL}) holds at small $x\lsim 10^{-2}$.
In order to model the sea contribution at larger $x$ we multiply (\ref{eq:F2BFKL})
by the familiar factor $(1-x)^m$, with $m=5$.
This factor does not affect the diffraction region but 
strongly suppresses production of  gluons with $x\gsim 0.1$.

 In applications it is convenient to work with $f_n (Q^2)$
represented in an analytical form. For the leading singularity we have 
\beq
f_0(Q^2)=
a_0{R_0^2Q^2\over{1+ R_0^2Q^2 }}
\left[1+c_0\log(1+r_0^2Q^2)\right]^{\gamma_0}\,,
\label{eq:F20}
\eeq
which has the large-$Q^2$ asymptotics \cite{PLNZ1,NZZJETP}
\beq
f_0(Q^2)\propto [\alpha_s(Q^2)]^{-\gamma_0} ,\,\,\, \gamma_0={4\over 3\Delta_0}\,.
\label{eq:F0}
\eeq

For $n\geq 1$ the functions $f_n(Q^2)$ can be approximated by
\beq
f_n(Q^2)=a_n f_0(Q^2){1+R_0^2Q^2\over{1+ R_n^2Q^2 }}
\prod ^{n_{max}}_{i=1}\left(1-{z\over z^{(i)}_n}\right)\,,
\label{eq:FN}
\eeq
where
\beq
z=\left[1+c_n\log(1+r_n^2Q^2)\right]^{\gamma_n}-1 ,\,\,\,
\gamma_n=\gamma_0 \delta_n 
\label{eq:ZFN}
\eeq
and $n_{max}=$min$\{n,2\}$.

Since the relevant variable is a power of the inverse gauge coupling
the nodes of $f_n(Q^2)$ are spaced by 2-3 orders of magnitude in $Q^2$-scale
 and only the first  two of them  are  in the accessible range of 
$Q^2$ \cite{DIS97, JETP}.
The first nodes of sub-leading $f_{n}(Q^2)$
are located at $Q^2\sim 20-60\, GeV^2$, the second nodes of $f_{2}(Q^2)$ and 
$f_{3}(Q^2)$ are at $Q^2\simeq 5\cdot 10^3\, GeV^2$ and 
$Q^2\simeq 2\cdot 10^4\, GeV^2$, respectively.
 The parameterization  tuned to reproduce
 the numerical results for $f_n(Q^2)$ at $Q^2\lsim 10^5\, GeV^2$
is given by eq.(\ref{eq:FN}). For $n=3$ we take a simplified form with only two
first nodes, because the third node of $f_3(Q^2)$ is at $\sim 2\cdot 10^7\, GeV^2$,
way beyond the reach of accelerator experiments at small $x$.
The found parameters are listed in the Table.\\

\vspace{0.5cm}

\begin{tabular}{|l|l|l|l|l|l|l|l|l|} \hline
$n$ & $a_n$  & $c_n$ & $r_n^2\,,$ ${\rm GeV^{-2}}$ &
$ R_n^2\,,$ ${\rm GeV^{-2}}$ &
$z^{(1)}_n$ & $z^{(2)}_n$ & $\delta_n$& $\Delta_n$ \\ \cline{1-9}
0 & 0.0232  & 0.3261&1.1204&2.6018& & & & 0.40 \\ \cline{1-9}
1 & 0.279 &0.1113&0.8755&3.4648&2.4773 &    &1.0915& 0.220\\ \cline{1-9}
2 & 0.195 &0.0833&1.5682&3.4824 &1.7706 &12.991 &1.2450& 0.148\\ \cline{1-9}
3 &0.471&0.0653&3.9567&2.7756    &1.4963 &6.9160  &1.2284& 0.111\\ \cline{1-9}
\end{tabular} 

\vspace{0.5cm}

Asymptotically, at $1/x\to\infty$, the expansion (\ref{eq:F2BFKL}) is dominated
by the term $f_0(Q^2)(x_0/x)^{\Delta_0}$. At moderately small $x$ the sub-leading
 terms are equally important since $\Delta_n\sim 1/n$. 
However, as it has been pointed out in \cite{DIS97,JETP}, 
for $Q^2\lsim 10^4\, {\rm GeV^2}$ all $f_n(Q^2)$
 with $n\geq 3$ are very close in shape to each other. Then we arrive at
the truncated expansion
\beq
F_2(x,Q^2)=\sum_{n=0}^3 f_n(Q^2)(x_0/x)^{\Delta_n}+
F_2^{soft}(Q^2)+F_2^{val}(x,Q^2)\,,
\label{eq:F2TRUN}
\eeq
where the term  $f_3(Q^2)(x_0/x)^{\Delta_3}$ with the properly
 adjusted weight factor, $a_3$,  stands for  all terms with $n\geq 3$.
 The addition of this ``background'' term  in  eq.(\ref{eq:F2TRUN})
 improves significantly the agreement with data for large $Q^2$ thus 
expanding the applicability region of eq.(\ref{eq:F2TRUN}) 
over the whole small-$x$ kinematical domain of  HERA.

The need for a soft pomeron contribution $F_2^{soft}$ in addition to the
perturbative gBFKL SF's   described previously is brought about by
phenomenological considerations.
A viable gBFKL phenomenology of the rising component of the proton
structure function over the whole range of $Q^{2}$ studied at
HERA (real photo-absorption included) is obtained if one starts
with the Born dipole cross section  $\sigma_{B}(r)$
as a boundary condition for the gBFKL evolution at $x_{0}=0.03$ \cite{NZHERA, JETP}.
However, such a purely perturbative input, $\sigma_{B}(r)$,
with $R_{c} = 0.27$\,fm strongly underestimates the cross sections of
soft processes and the proton SF at moderate $Q^2\sim 1\, {\rm GeV^2}$.
Therefore, at $r\gsim R_{c}$, the above described perturbative gBFKL
dipole cross section  $\sigma_{pt}(x,r)$, must be complemented
by the contribution from the non-perturbative soft pomeron,
$\sigma_{npt}(x,r)$. In terms of the relationship \cite{NZHERA} between $\sigma(x,r)$
and the gluon structure function of the proton, $G(x,Q^{2})$,
  the non-perturbative dipole cross section $\sigma_{npt}(r)$ at
$r\gsim R_c$ must be associated with soft non-perturbative 
 gluons in the conventional $G(x,Q^{2})$.
 The contribution to $G(x,Q^{2})$ from  the non-perturbative transverse momenta
$k^2\lsim Q_0^2\sim m^2_{\rho}$ persists at all $Q^2$ and must not be
subjected to the DGLAP evolution. 

Because the BFKL rise of  $\sigma(x,r)$ is due to production of 
$s$-channel perturbative gluons, which does not contribute to
$\sigma_{npt}(r)$ in \cite{NZHERA,JETP} we argued that to 
a first approximation one must consider the energy independent
$\sigma_{npt}(r)$  and additivity of scattering amplitudes from
both the hard BFKL and soft non-perturbative mechanisms. 
 For recent suggestions to identify our $\sigma_{npt}(r)$ 
with the soft pomeron of the two-pomeron picture see
\cite{LANDSH,GOLEC}. In  the models 
of soft scattering via polarization of the non-perturbative
QCD vacuum  \cite{Nachtmann,Dosch}, $\sigma_{npt}(r)$
is interpreted in terms of the  non-perturbative gluon distributions.

To our opinion, the recently encountered troubles with the small-$Q^2$
extrapolations of DGLAP evolution \cite {DICK} and the failure of DGLAP
fits in the Caldwell plot \cite{DERLOG}, \cite{DERLOG1}
 are due  to illegitimate enforcing
the DGLAP evolution upon the non-perturbative glue.

The non-perturbative term  $F_2^{soft}(Q^2)$ in eq.(\ref{eq:F2TRUN})
calculated  from eq.(\ref{eq:F2CC} ) with $\sigma=\sigma_{npt}(r)$
 from \cite{VECTOR} can be parameterized as follows
\beq
F_2^{soft}(Q^2)= 
b{R^2Q^2\over{1+ R^2Q^2 }}
\left[1+c\log(1+r^2Q^2)\right]\,,
\label{eq:F2NPT}
\eeq
where $b=0.1077$, $c=0.0673$, $R^2=6.6447\, {\rm GeV^{-2}}$ and 
$r^2=7.0332\, {\rm GeV^{-2}}$.
So, its $\log Q^2$-derivative levels off at very small $Q^2\sim 0.15\, {\rm GeV^2}$
and does not contribute to the observed growth of $\partial F_2/\partial\log Q^2$.

In Fig.1 we confront our estimates to both the HERA data and the
fixed target data. In \cite{DERLOG1} the logarithmic slope,
 $\partial F_2/\partial\log Q^2$, is derived from data by fitting
$F_2=a+b\log Q^2$ in bins of fixed $x$. The average value of $Q^2$, 
$\langle Q^2\rangle$, in each $x$-bin
 is derived from the $F_2$ weighted mean value of $\log Q^2$ in that bin.

As we have noticed above, at moderately small $x\sim 10^{-2}-10^{-3}$  
the contribution of the sub-leading poles to  $F_2(x,Q^2)$ is still substantial
(the relative weight factors, $a_n$, with $n\geq 1$ are presented in the table),  
 but toward the region of $x\sim 10^{-6}$ the leading pole contribution starts
to prevail.At small $Q^2$ the ratio of $\log Q^2$-derivatives,
 $r_n=f^{\prime}_n/f^{\prime}_0$, can be estimated as 

\beq
r_n\simeq a_n{\Lambda^2_0\over \Lambda^2_n}\left(x\over x_0\right)^{\Delta_0-\Delta_n} 
\label{RATIO}
\eeq
where 

\beq
\Lambda_n^2 \simeq 1/(R^2_n-c_0\gamma_0r_0^2)\,.
\label{eq:LAMB}
\eeq

Because the  sub-leading SF's, $f_n(Q^2)$,
 have node at $Q^2\sim  20-60\,{\rm GeV^2}$ \cite{DIS97,JETP},
 their contribution to the slope  $\partial F_2/\partial\log Q^2$ 
vanishes at $Q^2\sim 5-10 \, {\rm GeV^2}$, which is very close to 
 the turn-over point in the HERA data.
Hence  $\partial F_2/\partial\log Q^2$ at small 
$Q^2$ follows closely $\partial f_{0}/\partial\log Q^2$.
From (\ref{eq:F20}) it follows that  at small $Q^2$, $f_{0}(Q^2)$ behaves like
 $\sim Q^2/(\Lambda_0^2+Q^2)$ with $\Lambda_0^2\simeq 0.72\,{\rm GeV^2}$ 
coming from (\ref{eq:LAMB}). Therefore, $\partial F_{2}/\partial\log Q^2$ 
rises with $Q^2$ up to $Q^2\sim 1 {\rm GeV^2}$ then levels off.
Only at large $Q^2$, when the sub-leading terms enter the game,
$\partial F_2/\partial\log Q^2$ decreases and even becomes negative valued at
  large $x$. Our estimates shown in Fig.1a are in good agreement 
with HERA data \cite{DERLOG1}. The curves are somewhat wiggly because
the $x-\langle Q^2\rangle$ correlation of the experimental data  is 
non-monotonous one.  
 
 In Fig.1b we compare our predictions with the fixed target data \cite{DERLOG}. 
Variation of the slope in this case is less pronounced since the starting value of 
$\langle Q^2\rangle$ is $\langle Q^2\rangle \simeq 0.54\,{\rm GeV^2}$ 
at $x\simeq 10^{-3}$ (compare with $Q^2=0.12\,{\rm GeV^2}$ 
at $x=2.1\cdot 10^{-6}$ at HERA). It can easily be seen that the derivative 
$\partial f_{0}/\partial\log Q^2$ at such $Q^2$ is  a rather slow function of $Q^2$.
The agreement of our estimates with the fixed target data is quite reasonable,
though  there is a systematic discrepancy   at small $x$. 
We recall that there is a certain mismatch between the E665 and H1/ZEUS data
on $F_2(x,Q^2)$ in the close $(x, Q^2)$ bins (see Fig.2a and Fig.2b).


\newpage

{\large\bf{ Figure Captions}\\ }
\begin{enumerate}
\item[{\bf Fig.1}]
Caldwell's plot of $\partial F_2/\partial \log Q^2$ for the ZEUS data 
\cite{DERLOG1} (Fig. 1a)
 and fixed target data \cite{DERLOG} (Fig. 1b).
 Our  predictions (BFKL-Regge) are shown by 
the  solid lines.  Shown by the dashed lines is the leading BFKL pole
approximation (LPA). 

\item[{\bf Fig.2}]
Description of the H1, ZEUS and E665 $F_2(x,Q^2)$ data 
by the BFKL-Regge expansion
(\ref{eq:F2BFKL}): the large-$Q^2$ data
 ($Q^2=$3.5, 12, 25, 65, 120 and 200 GeV$^2$)
 are shown in Fig.2a, 
the small-$Q^2$ data ($Q^2=$0.11, 0.20, 0.40, 0.65, 0.85 and 1.2 GeV$^2$) are
in Fig.2b. For display purposes we have multiplied $F_2$ by the
 numbers shown in brackets. 
\end{enumerate}
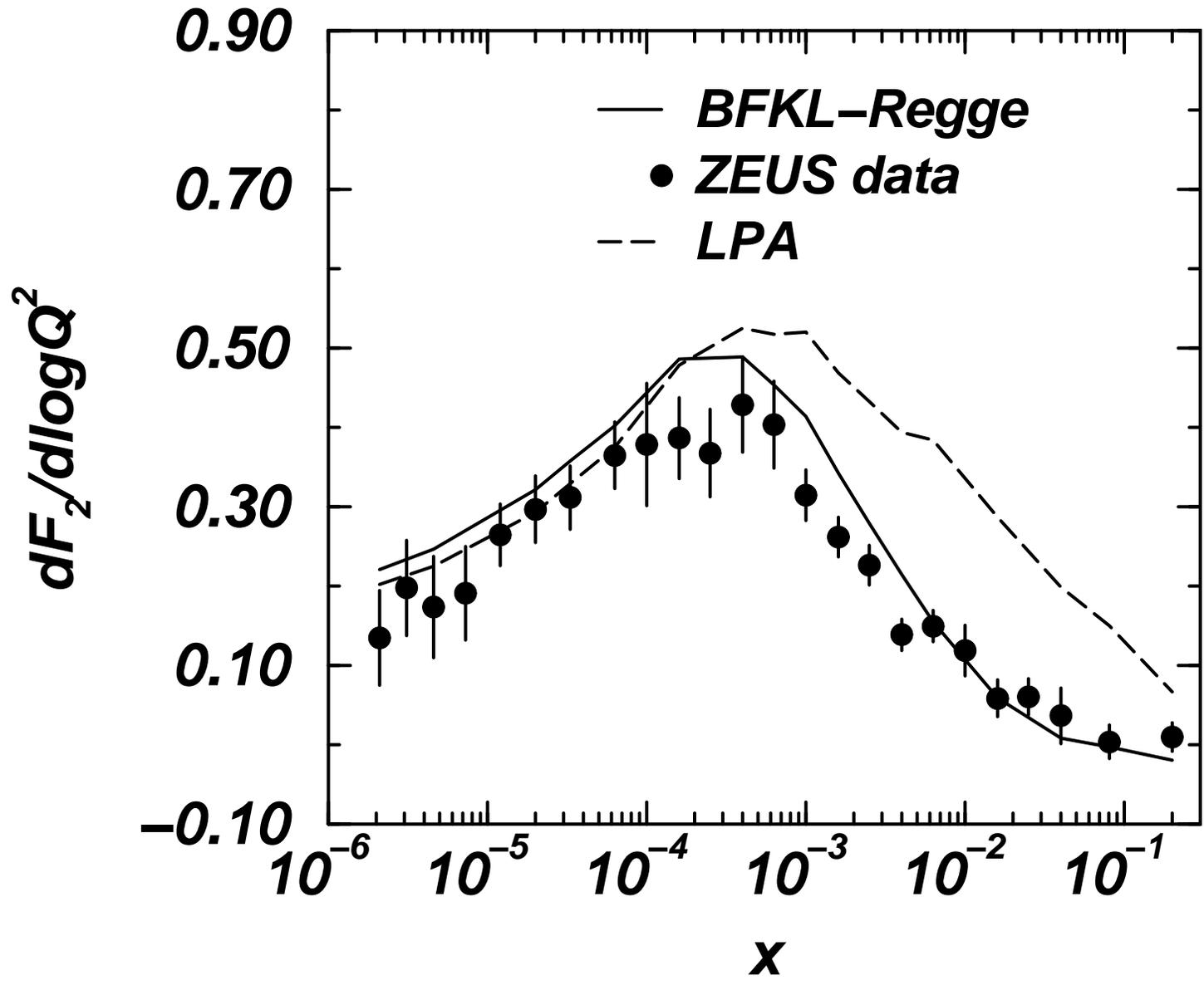
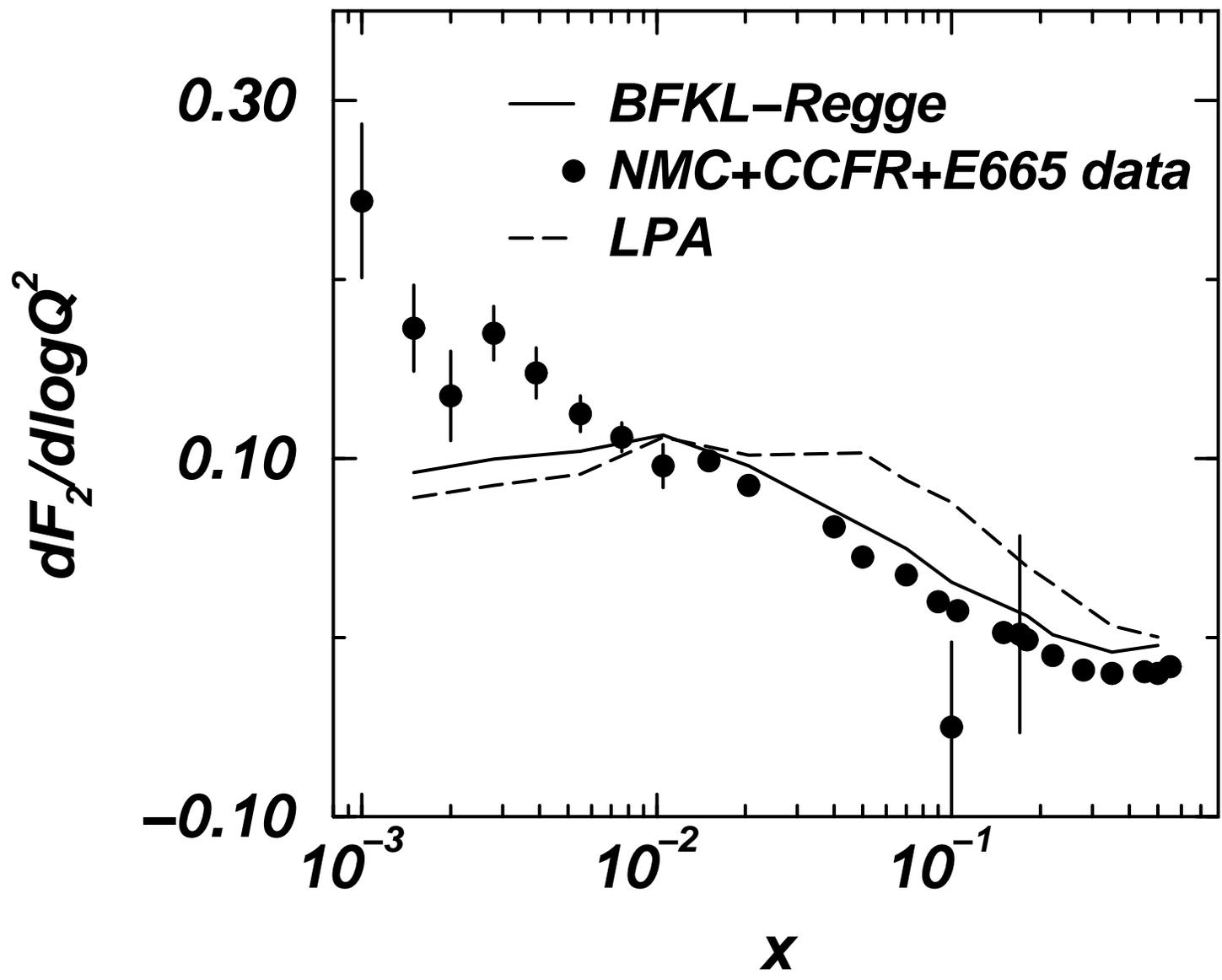
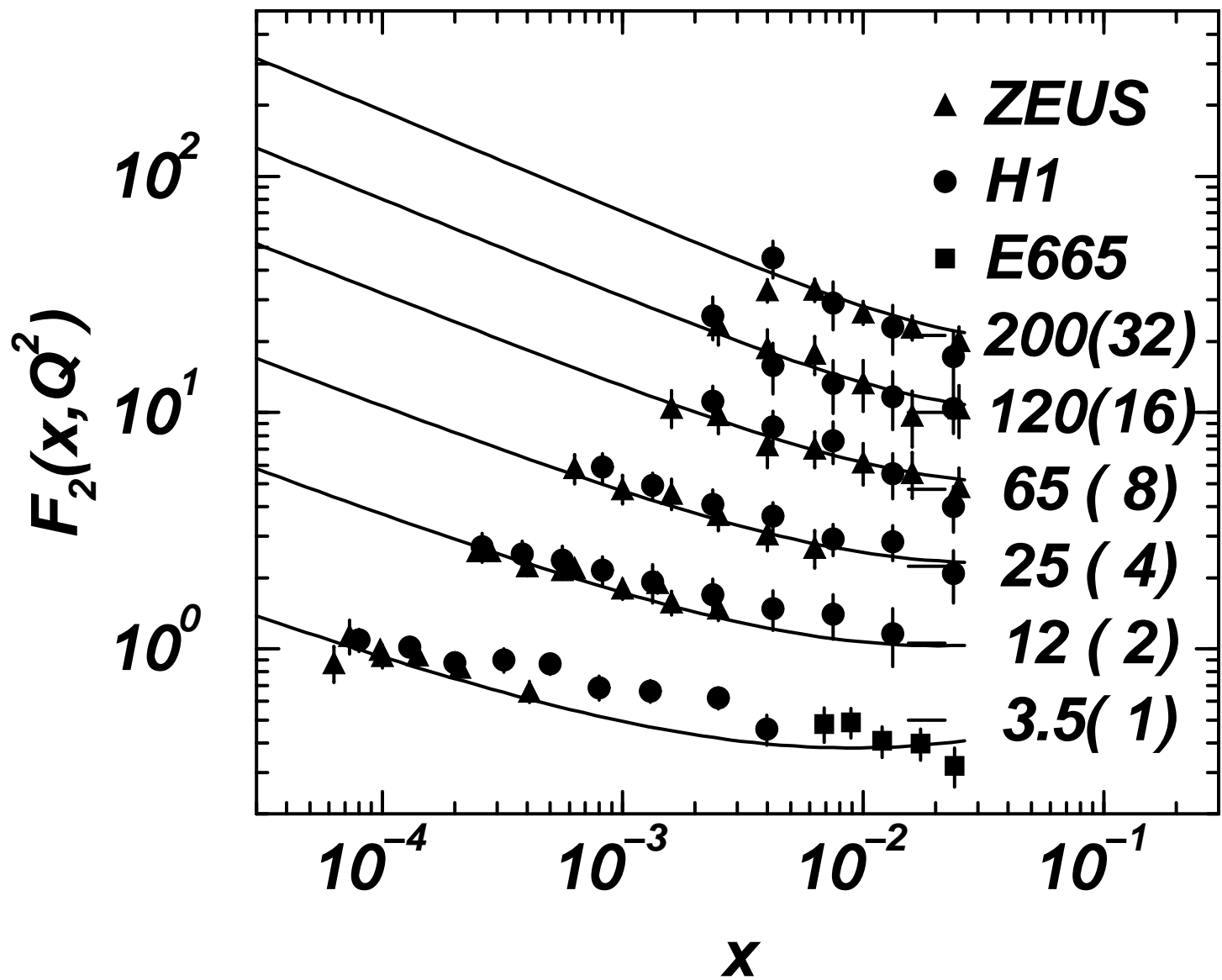
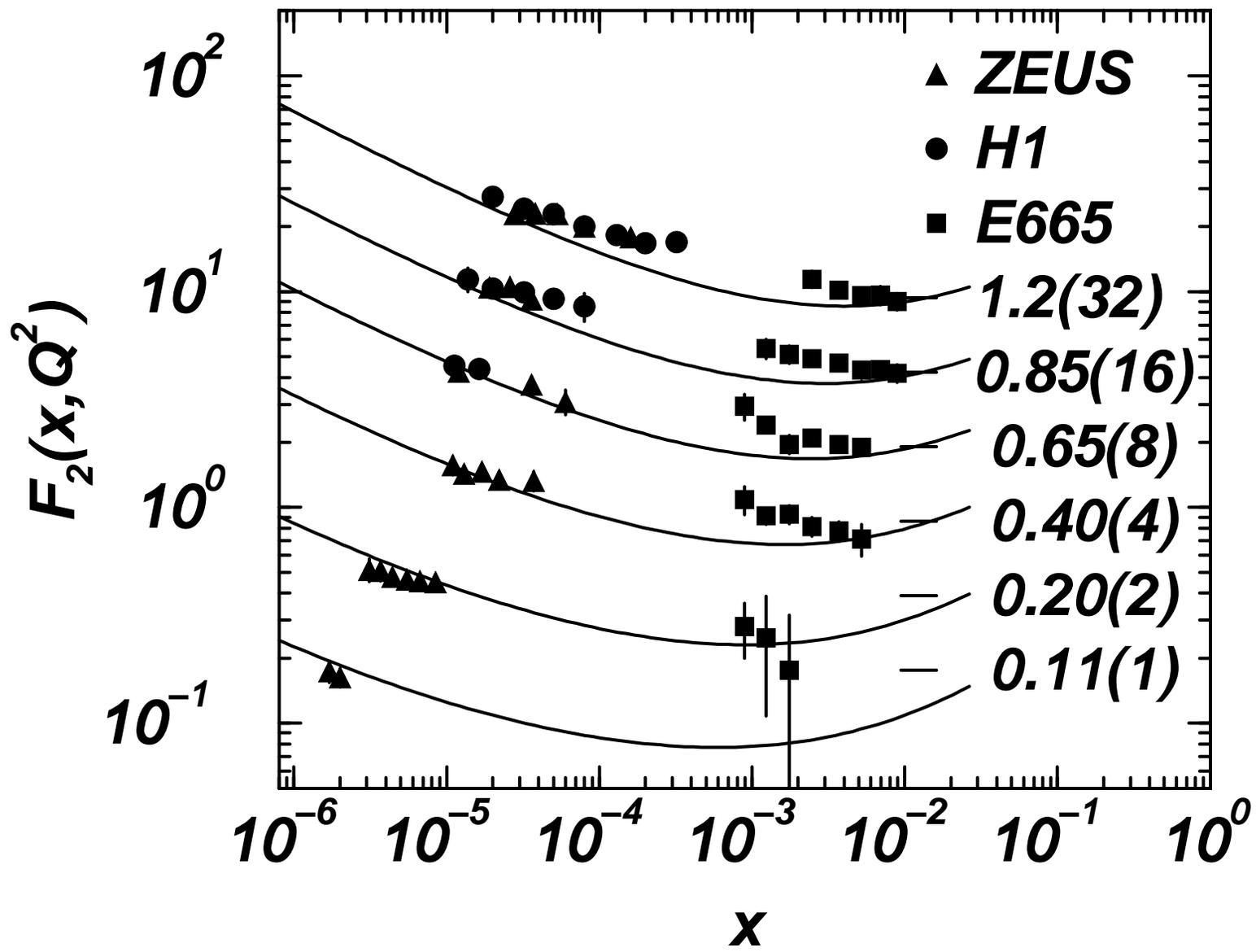
\end{document}